\documentclass[runningheads]{llncs}

\usepackage{graphicx}
\usepackage{colortbl}
\usepackage{caption}
\usepackage{subcaption}

\usepackage{hyperref}
\hypersetup{pdfborder=0 0 0}
\usepackage{cleveref}
\usepackage{orcidlink}

\usepackage{booktabs}
\usepackage{tabularx}
\usepackage{colortbl}

\definecolor{archived-blue}{rgb}{0,0,1}
\definecolor{os-green}{rgb}{0.416, 0.66, 0.31}
\definecolor{reachable-yellow}{rgb}{0.945,0.76,0.196}
\definecolor{uponrequest-brown}{rgb}{0.75,0.565,0}
\definecolor{broken-orange}{rgb}{0.9, 0.57, 0.22}
\definecolor{unavailable-red}{rgb}{0.6,0,0}
\definecolor{private-grey}{rgb}{0.72,0.72,0.72}
\definecolor{proprietary-grey}{rgb}{0.26,0.26,0.26}

\begin{document}

\title{NLP4RE Tools: Classification, Overview, and Management}

\author{
    Julian Frattini\inst{1}\orcidlink{0000-0003-3995-6125} \and
    Michael Unterkalmsteiner\inst{1}\orcidlink{0000-0003-4118-0952} \and
    Davide Fucci\inst{1}\orcidlink{0000-0002-0679-4361} \and
    Daniel Mendez\inst{1,2}\orcidlink{0000-0003-0619-6027}
}
\authorrunning{J. Frattini et al.}
\institute{
    Blekinge Institute of Technology, Valhallavägen 1, 371 41 Karlskrona, Sweden \email{\{firstname\}.\{lastname\}@bth.se} \and
    fortiss GmbH, Guerickestraße 25, 80805 Munich, Germany \email{daniel.mendez@fortiss.de}
}

\maketitle

\begin{abstract}
    Tools constitute an essential contribution to natural language processing for requirements engineering (NLP4RE) research.
    They are executable instruments that make research usable and applicable in practice. 
    In this chapter, we first introduce a systematic classification of NLP4RE tools to improve the understanding of their types and properties.
    Then, we extend an existing overview with a systematic summary of 126 NLP4RE tools published between April 2019 and June 2023 to ease reuse and evolution of existing tools.
    Finally, we provide instructions on how to create, maintain, and disseminate NLP4RE tools to support a more rigorous management and dissemination.
\end{abstract}

\section{Introduction}
\label{sec:intro}

The prevalence of unrestricted natural language (NL) to specify requirements in the requirements engineering (RE) phase of a software development project~\cite{franch2023state} complicates approaches for automation due to the inherent complexity and ambiguity of NL.
Natural language processing (NLP) techniques meet this challenge through the application of machine and deep learning techniques to NL~\cite{eisenstein2019nlp}.
An essential contribution to this natural language processing for requirements engineering (NLP4RE) research domain are NLP4RE \textit{tools}. 
They constitute a primary research artifact to transfer technology into practice~\cite{gorschek2006model} by automating a variety of tasks relevant to RE or by providing decision support to managers in practice.
These tasks include requirements elicitation, e.g., by mining new requirements from a large body of user reviews~\cite{wang2020opinion} or by extracting relevant clauses from regulatory text~\cite{anish2021automated}, but also requirements management, i.e., retrieving dependencies between requirements among each other~\cite{deshpande2020requirements} or between requirements and other software artifacts~\cite{unterkalmsteiner2020tt}, requirements modeling, e.g., the transformation of NL requirements into cause-effect graphs~\cite{frattini2020automatic} or state models~\cite{madala2021model}, and validation and verification, like automatically or semi-automatically generating test cases from NL requirements~\cite{frattini2023cira}.
The degree of support ranges from attempts at full automation~\cite{fischbach2023automatic,frattini2023cira} to recommendations aiming to ease manual tasks~\cite{unterkalmsteiner2020tt}.

However, developing, deploying, maintaining and disseminating NLP4RE tools is challenging and sometimes even unnecessary, as similar tools may already exist, yet overlooked due to the lack of a comprehensive overview.
To meet these challenges and equip the reader with the knowledge to contribute to the landscape of NLP4RE tools, this chapter treats the following three topics:

\begin{itemize}
    \item \textbf{Classification} (\Cref{sec:classification}): a systematic classification of NLP4RE tools to categorize them by relevant attributes.
    \item \textbf{Overview} (\Cref{sec:overview}): a comprehensive overview of existing NLP4RE tools to show which tools are already available and can be reused instead of re-implementing already existing solutions.
    \item \textbf{Management Guidelines} (\Cref{sec:management}): instructions on developing, documenting, and disseminating NLP4RE tools to make contributing to the landscape of existing tools more approachable.
\end{itemize}

\section{Classification}
\label{sec:classification}

This section introduces a classification of NLP4RE tools.
This classification allows for a systematic understanding of the types of tools and the attributes in which they differ, but also aids in reporting new NLP4RE tools.
The classification is general enough to be applicable to any RE tool, not just those processing NL requirements, which allows future extensions to a larger scope.
Adherence to this general classification scheme is a first and accessible step towards a more consistent reporting of NLP4RE tools, and it paves the way for future, more comprehensive approaches~\cite{abualhaija2023replication}.

The classification is derived from a previous NLP4RE tool classification scheme~\cite{zhao2021natural}, and complemented by an ontology of requirements quality factors~\cite{frattini2022live}.
The tool classification scheme stems from a seminal, comprehensive secondary study on NLP for RE~\cite{zhao2021natural}.
The ontology extends several aspects of this scheme, especially regarding the type and availability of the tool~\cite{frattini2022live}, which improves the overview of the tools' accessibility.

We focus on the attributes listed in \Cref{tab:attributes}.
\Cref{sec:classification:name} presents the attributes describing the \textbf{identification} of the tool, i.e., their \textit{name} and \textit{description}.
\Cref{sec:classification:availability} details attributes of the \textbf{artifact} itself, i.e., the link to the \textit{source}, its \textit{release} type, \textit{availability}, and \textit{license}.
\Cref{sec:classification:task} specifies the \textbf{task} which the tool performs in terms of supported \textit{RE activity} and \textit{task type}.
This selection of attributes suffices, in our experience, to categorize NLP4RE tools at a high level and it serves the purpose of providing an overview of NLP4RE tools. 
For a more detailed classification, the interested reader may refer to the \texttt{NLP4RE ID card} by Abualhaija et al.~\cite{abualhaija2023replication}.
\Cref{tab:examples} lists the classification of two tools~\cite{fischbach2021cira,unterkalmsteiner2020tt} which will be used as running examples in the following subsections.

\begin{table}[hbt]
    \centering
    \caption{Attributes of NLP4RE tools}
    \label{tab:attributes}
    \begin{tabularx}{\textwidth}{l|l|X}
        \toprule
        \textbf{Attribute} & \textbf{Type} & \textbf{Values} \\ \midrule
        \multicolumn{3}{l}{\textbf{Identification}} \\ \hline
        Name & text & \\ 
        Description & text & \\ \hline
        \multicolumn{3}{l}{\textbf{Artifact}} \\ \hline
        Source & text & URL \\
        Release & categorical & \{source code, executable\} \\
        Availability & categorical & \{archived, open source, reachable, upon request, broken, unavailable, private, proprietary\} \\
        License & categorical & \{MIT, Apache 2.0, GPL, ... \} \\ \hline
        \multicolumn{3}{l}{\textbf{Task}} \\ \hline
        RE Activity & categorical & \{elicitation, analysis, modeling, validation \& verification, management, other\}\\
        Task type & categorical & \{modeling, detection, extraction, classification, tracing \& relating, search \& retrieval\} \\ 
        \bottomrule
    \end{tabularx}
\end{table}

\begin{table}[hbt!]
    \caption{Characterization of the CiRA~\cite{fischbach2021cira} and TT-RecS tools~\cite{unterkalmsteiner2020tt}}
    \label{tab:examples}
    \centering
    \begin{tabularx}{\textwidth}{l|XX}
        \toprule
        \multicolumn{3}{l}{\textbf{Identification}} \\ \hline
        Name & \textbf{CiRA} & \textbf{TT-RecS} \\
        Description & A tool that classifies a natural language requirements sentence as either causal or non-causal by leveraging a fine-tuned BERT model & The tool supports the annotation of requirements with concepts from a domain-specific taxonomy, providing suggestions which concepts fit to the requirement content. The goal is to enable traceability through the use of a taxonomy, assuming that downstream artifacts are also classified. \\ \hline
        \multicolumn{3}{l}{\textbf{Artifact}} \\ \hline
        Source & \url{https://zenodo.org/record/8033347} & \url{https://zenodo.org/records/3827169} \\
        Release & Source code & Source code \\
        Availability & Archived & Archived \\
        License & Apache 2.0 & Apache 2.0 \\ \hline
        \multicolumn{3}{l}{\textbf{Task}} \\ \hline
        RE Activity & Validation \& Verification & Management \\
        Task type & Classification & Tracing \& Relating \\
        \bottomrule
    \end{tabularx}
\end{table}

\subsection{Identification}
\label{sec:classification:name}

Every NLP4RE tool should be characterized by a descriptive \textit{name} and a comprehensive \textit{description}, though many tools do not bear a distinct name. 
There are no restrictions to tool names, but one popular pattern is to name a tool after the abbreviation of a concise description of its objective, e.g., the CiRA classifier, a \textbf{c}ausality \textbf{i}n \textbf{r}equirements \textbf{a}rtifacts classifier~\cite{fischbach2021cira}, or TT-RecS, a \textbf{t}axonomic \textbf{t}race \textbf{rec}ommender \textbf{s}ystem~\cite{unterkalmsteiner2020tt}.

The description of a tool should fulfill two purposes: 
(1) it should describe the main objective of the tool, and (2) it should explain the key technologies used.
The first purpose is tailored towards users of the tool who are only interested in a black-box view of its functionality.
The second purpose pertains towards researchers and developers who want to assess the tool's technological components and assess whether and how they could improve it.
For example, the descriptions of the two tools provided in \Cref{tab:examples} contain black-box input and output combination as well as information about the tools' core technology.
While scientific manuscripts---other than technical reports---often constrain these descriptions to a very high level, this purpose can be achieved to a much greater degree when properly documenting the tool by describing the processing pipeline, adding architectural views, and specifying the core algorithms using pseudo code.
The repository hosting the tool described in a manuscript is the appropriate location for extending these technological descriptions, as further explained in \Cref{sec:management:development:doc}.

\subsection{Artifact}
\label{sec:classification:availability}


The source link is a URL pointing to the location of the tool, like \url{https://zenodo.org/record/8033347} for the CiRA classifier~\cite{fischbach2021cira}.
The type of release denotes what kind of artifact has been made available.
For tools, the type of release is usually either \textit{source code} or a pre-compiled \textit{executable}, but in rare cases authors release both.
The degree to which the tool can be accessed constitutes the availability of a tool. 
\Cref{tab:availability} shows an eight-tier ranking of availability levels proposed in prior research~\cite{frattini2023stop}.

\begin{table}[hbt!]
    \centering
    \footnotesize
    \caption{Levels of availability~\cite{frattini2023stop} }
    \label{tab:availability}
    \begin{tabularx}{\textwidth}{l|X} \toprule
        \textbf{Status} & \textbf{Explanation} \\ \hline
        \cellcolor{archived-blue} {\color{yellow} Archived} & The tool is hosted in a service that satisfies the following criteria: (1) \textbf{immutable URL} (cannot be altered by the author or someone else), (2) \textbf{permanent} (the hosting organization has a mission to maintain artifacts for the foreseeable future), and (3) \textbf{accessible} (there is a DOI pointing to the real data source URL) \\
        \cellcolor{os-green} Open Source & The source code is disclosed and contains an opensource license which grants access and re-use \\
        \cellcolor{reachable-yellow} Reachable & The tool is reachable now but missing some of the \textit{archived} or \textit{open source} aspects\\
        \cellcolor{uponrequest-brown} Upon request & Authors claim the tool is available upon request \\ \hline
        \cellcolor{broken-orange} Broken & A link to the tool is contained in the paper, but it does not resolve\\
        \cellcolor{unavailable-red} {\color{white} Unavailable} & A tool is presented, but no indication on how to access it is provided \\
        \cellcolor{private-grey} Private & The authors state that a tool exists, but it is private for some reasons (such as industry collaboration with private data, etc.) \\
        \cellcolor{proprietary-grey} {\color{white} Proprietary} & The tool is proprietary, and access is granted upon payment \\
        \bottomrule
    \end{tabularx}
\end{table}

In the best case, tools are \textit{archived} properly, which ensures their long-term availability.
In most cases, however, tools are simply \textit{unavailable} without explanation.
This significantly inhibits adoption, replication, and evolution of contributions to the NLP4RE research domain~\cite{frattini2023stop}.
\Cref{sec:overview} presents the current distribution of tool availability and \Cref{sec:management} guidelines to ensure availability.

The fourth attribute of the artifact category, the license, has a major influence on its availability.
A license specifies if and how a third party can use the artifact and is, hence, imperative for the reuse and evolution of NLP4RE tools.
\Cref{sec:management:dissemination:licensing} describes the types and properties of licenses in more detail.

\subsection{Task}
\label{sec:classification:task}

Finally, Zhao et al.~\cite{zhao2021natural} classify NLP4RE tools based on the \textit{RE activity}, which the tool intends to support, e.g., elicitation or analysis, and the \textit{task} type, which the tool employs, e.g., classification or extraction.
These two attributes are orthogonal to each other, i.e., different tasks can be used to facilitate different RE activities.
The RE activity attribute denotes the high-level requirements engineering activity that the tool supports. 
This attribute contains the following categories adapted from Zhao et al.~\cite{zhao2021natural}:

\begin{itemize}
    \item Elicitation: Emergence of new requirements
    \item Analysis: Assessment of requirements specifications through attribution and evaluation
    \item Modeling: Translation of requirements into different representations (e.g., UML models or code)
    \item Validation \& Verification: Assertion of whether the behavior of a system under test matches the expected behavior as specified in requirements
    \item Management: Coordination and evolution of requirements and their interrelations
    \item Multiple: Several of the above
    \item Other: Any RE-activity not covered by the aforementioned
\end{itemize}

As specified in \Cref{tab:examples}, the CiRA classifier supports the RE activity validation \& verification, as the classification serves to identify causal requirements which can be automatically translated into test cases~\cite{fischbach2023automatic}.
The TT-RecS tool supports the RE activity management, as it recovers trace links between requirements and other software requirements~\cite{unterkalmsteiner2020tt}.

The task type denotes the NLP task which the tool is fulfilling.
Where the previously explained RE activity determines \textit{what} a tool is supposed to do, the task type defines \textit{how} it achieves this goal.
Note that an NLP4RE tool can employ different task types to achieve a similar goal.
The task type attribute contains the following categories adapted from Zhao et al.~\cite{zhao2021natural}:

\begin{itemize}
    \item Generation: Formation of new, requirements-relevant text
    \item Detection: Detecting linguistic issues
    \item Extraction: Identifying key domain abstractions and concepts
    \item Classification: Classifying NL text into different categories
    \item Transformation: Translation of text into a semantically similar but syntactically different specification
    \item Tracing \& Relating: Establishing a relationship between requirements and/or other software artifacts
    \item Search \& Retrieval: Searching and retrieving requirements-relevant information from existing corpora
\end{itemize}

For example, the CiRA tool fulfills a \textit{classification} task to distinguish causal from non-causal requirements sentences.
The TT-RecS tool fulfills a \textit{tracing \& relating} task to recover trace links between requirements and other software artifacts.
The two attributes presented in this section, RE activity and task type, constitute the essence of an NLP4RE tool:
The task type represents the NLP part and the RE activity the RE part of the tool.
The previously introduced \textit{description} attribute can easily be constructed from these two attributes.

\section{Overview}
\label{sec:overview}

The availability of NLP technology has given rise to a large number of NLP4RE tools~\cite{zhao2021natural}.
The abundance of available tools makes it difficult to maintain an overview, causing too often overlapping and competing research endeavors~\cite{frattini2022live}.
Consequently, systematic studies updating an overview and categorization of existing tools are necessary.
This chapter continues the overview over NLP4RE tools initiated by Zhao et al.~\cite{zhao2021natural}.
\Cref{sec:overview:method} explains the method of extraction, \Cref{sec:overview:result} presents the obtained results, and \Cref{sec:overview:discussion} discusses them.

\subsection{Method}
\label{sec:overview:method}

Zhao et al. contributed a first, systematic overview of the NLP4RE landscape in their seminal mapping study~\cite{zhao2021natural}.
This study also included an investigation of NLP4RE tools up until April 2019.
We continued this systematic study of NLP4RE tools.
All study material can be found in our replication package.\footnote{\url{https://zenodo.org/doi/10.5281/zenodo.8341439}}

\subsubsection{Article Selection}
\label{sec:overview:method:selection}

As data sources, we selected the twelve leading publication venues for NLP4RE research as identified by Zhao et al.~\cite{zhao2021natural} listed in the left-most column of \Cref{tab:articles:initial}.
This list included RE- and NLP-specific venues like the Requirements Engineering Conference\footnote{\url{https://requirements-engineering.org/}} and the NLP4RE workshop\footnote{\url{https://nlp4re.github.io/2023/}} but also general software engineering venues like the Journal of Systems and Software\footnote{\url{https://www.sciencedirect.com/journal/journal-of-systems-and-software}} and the Transactions in Software Engineering journal.\footnote{\url{https://www.computer.org/csdl/journal/ts}}

\begin{table}[hbt!]
    \centering
    \caption{Number of included articles (journals, conferences, and workshops)}
    \label{tab:articles:initial}
    \begin{tabularx}{\textwidth}{Xl|rrrrr|r}
        \toprule
        \textbf{Venue} & \textbf{Acro.} & \textbf{2019} & \textbf{2020} & \textbf{2021} & \textbf{2022} & \textbf{2023} & \textbf{Total} \\ \midrule
        Transactions on Software Engineering & TSE & 0 & 1 & 0 & 1 & 2 & 4 \\
        Journal of Systems and Software & JSS & 0	& 2	& 0 & 1 & 1 & 4 \\
        Information and Software Technology & IST & 3 & 1 & 3 & 2 & 2 & 11 \\
        Requirements Engineering Journal & REJ & 1 & 1 & 2 & 2 & 0 & 6 \\
        Data \& Knowledge Engineering Journal & DKE & 0 & 2 & 0 & 0 & 0 & 2 \\ \hline
        International Conference on Software Engineering & ICSE & 0 & 0 & 1 & 4 & 5 & 10 \\
        Automated Software Engineering & ASE & 0 & 3 & 0 & 2 & 0 & 5 \\
        Requirements Engineering Conference & RE & 13 & 12 & 13 & 7 & 0 & 45 \\
        Requirements Engineering: Foundations of Software Quality & REFSQ & 2 & 0 & 1 & 3 & 2 & 8 \\
        Symposium on Applied Computing & SAC & 0 & 1 & 3 & 0 & 0 & 4 \\ \hline
        Artificial Intelligence for Requirements Engineering & AIRE & 3 & 6 & 6 & 2 & 0 & 17 \\
        Natural Language Processing for Requirements Engineering & NLP4RE & 1 & 1 & 5 & 2 & 1 & 10 \\ \hline
        \textbf{Total} & & 23 & 30 & 34 & 26 & 13 & 126 \\
        \bottomrule
    \end{tabularx}
\end{table}

The first three authors reviewed all publications in these twelve venues from April 2019 until June 2023, as far as available up to the time of conducting this study, by reading the title, abstract, and introduction.
For each article, we determined its potential relevance by deciding the two initial inclusion criteria PIn1 and PIn2 listed in \Cref{tab:criteria}.
Initial inclusion criterion PIn1 was necessary to filter the generic venues for RE-relevant articles.
Initial inclusion criterion PIn2 was necessary to filter for articles potentially presenting tools.
The initial inclusion produced 202 eligible articles.
After the selection of candidate articles, we conducted a proper inclusion and exclusion phase with the criteria In1-In3 and Ex1-Ex3 listed in \Cref{tab:criteria}.
After the inclusion and exclusion phase, 126 relevant articles remained.
\Cref{tab:articles:initial} lists the number of included articles per venue per year.

\begin{table}[hbt!]
    \centering
    \caption{Preliminary inclusion (PIn), final Inclusion (In) and exclusion (Ex) criteria for candidate articles}
    \label{tab:criteria}
    \begin{tabularx}{\textwidth}{l|X}
        \toprule
        \textbf{Index} & \textbf{Criterion} \\ \midrule
        PIn1 & The article is either published in a RE-centric venue or the title, abstract, or introduction explicitly mention the area of RE. \\
        PIn2 & The title, abstract, or introduction mention any sort of software or automation hinting at a tool. \\ \hline
        In1 & The article explicitly declares the relevance of the presented results to the area of requirements engineering by making clear which RE activity the tool supports. \\
        In2 & The article presents an automatic or semi-automatic algorithm and is an original contribution of the paper. \\
        In3 & The algorithm explicitly reports to involve at least one technology processing natural language, either as input or output, which is not limited to statistical NLP. \\ \hline
        Ex1 & The article is not accessible freely or through institutional access programs. \\
        Ex2 & The article is not written in English. \\
        Ex3 & The article is not extended by or a duplicate of an already included article. \\
        \bottomrule
    \end{tabularx}
\end{table}

\subsubsection{Data Extraction and Classification}
\label{sec:overview:method:extraction}

We prepared guidelines for the data extraction and classification phase, which can also be found in our replication package.
The extraction guidelines contained one section for each category mentioned in \Cref{sec:classification}.
Each section of the extraction guidelines consisted of a description of the attribute, the purpose of its extraction, the extraction rules to apply, and examples.
We distributed the extraction and classification of candidate articles among the first three authors based on their availability.
Each author performed the extraction and classification individually but took note of any difficulty encountered during the process.
These difficulties were discussed and resolved jointly among the three authors in a final meeting.

\subsubsection{Analysis and Dissemination}
\label{sec:overview:method:analysis}

Finally, we analyzed the extracted data by generating descriptive statistics of the distribution of the extracted attributes.
Additionally, we disseminated the results by creating a curated list on GitHub.\footnote{\label{foot:awesome-list}\url{https://github.com/JulianFrattini/nlp4re-tools/blob/main/tools/nlp4re-tools.md}}
With this list, we provide an up-to-date overview of the NLP4RE tools landscape.
The overview shall support researchers in finding and accessing available NLP4RE tools.
Furthermore, we invite fellow researchers to contribute to the list using GitHub's collaboration features in order to keep it up-to-date.

\subsection{Results}
\label{sec:overview:result}

\subsubsection{NLP4RE Tools}

\begin{figure}[hbt!]
    \centering
    \includegraphics[width=0.8\textwidth]{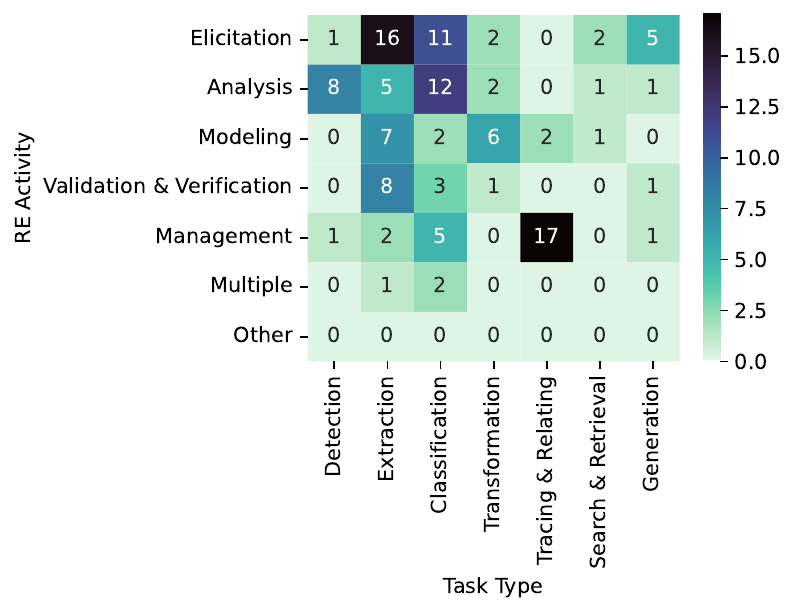}
    \caption{Distribution of RE activity and task type among tools}
    \label{fig:tooltypes}
\end{figure}

126 of the 202 candidate articles contained an NLP4RE tool according to our inclusion and exclusion criteria.
The full list of tools with all extracted attributes can be found online.\textsuperscript{\ref{foot:awesome-list}}
\Cref{fig:tooltypes} shows the distribution the RE activity and tool task codes among the extracted tools.

The distribution highlights several focal points of research presenting NLP4RE tools.
The requirements elicitation activity is most prominently supported by extraction and classification tools.
Classification tools for requirements elicitation focus on identifying information that is potentially relevant to the elicitation of new requirements from textual corpora, e.g., user reviews on app stores.
Techniques like deep learning~\cite{mekala2021classifying} and transfer learning~\cite{henao2021transfer} are applied to classify reviews into relevant (e.g., feature request) and irrelevant (e.g., noise) groups.
On the other hand, extraction tools for requirements elicitation fully automate the elicitation process by also consolidating the resulting set of relevant data into requirement proposals.
Tools like OpenReq Analytics~\cite{stanik2019requirements} or RE-BERT~\cite{de2021re} extract new requirements from app reviews, but there are also examples of tools considering different sources, like MaRK-II~\cite{lian2020assisting}, which processes domain documents.

The requirements analysis activity is mostly supported by detection, extraction, and classification tools.
Detection tools for requirements analysis identify potential requirements quality defects and include tools like the automatic ambiguity detector~\cite{ezzini2022automated} and MAANA~\cite{ezzini2021using}.
Extraction tools for requirements analysis support information retrieval tasks like REVV-light for near-synonymy detection~\cite{dalpiaz2019detecting} or Co-AI for retrieving abstractions~\cite{peng2021co}, i.e., significant domain terms.
Most prominently, classification tools for requirements analysis enrich requirements with additional information like a categorization of an NL requirements as functional or non-functional~\cite{hey2020norbert} or the distinction of requirements from prose in large specifications~\cite{falkner2019identifying}.

The requirements modeling activity is mainly supported by extraction and transformation tools.
Extraction tools for requirements modeling focus on retrieval of relevant concepts from natural requirements like i-Star elements~\cite{xiong2022bistar} or system states~\cite{pudlitz2019extraction}.
Transformation tools further automate the process by proposing actual models from NL requirements, for example full i-Star models~\cite{chen2019t} or cause-effect diagrams~\cite{frattini2020automatic}.

The validation and verification task is mostly supported by extraction tools.
These usually attempt to generate test cases from NL requirements by extracting causal diagrams~\cite{fischbach2019automated} and inferring test configurations from them~\cite{jadallah2021cate}.
Classification tools to identify NL requirements containing causal information, like the CiRA tool~\cite{fischbach2021cira} mentioned in \Cref{sec:classification}, further aid this process.

The management activity is predominantly aided by tracing and relating tools.
These tools recover trace links between requirements themselves~\cite{motger2019openreq} or between requirements and other SE artifacts~\cite{unterkalmsteiner2020tt}.
Only very few tools support multiple RE activities. 
Among these are, for example, very general algorithms like zero-shot learning for a variety of requirements classification tasks~\cite{alhoshan2023zero}.

\subsubsection{Artifact Availability}

\begin{figure}[hbt!]
    \centering
    \begin{subfigure}[T]{0.49\textwidth}
        \includegraphics[width=\textwidth]{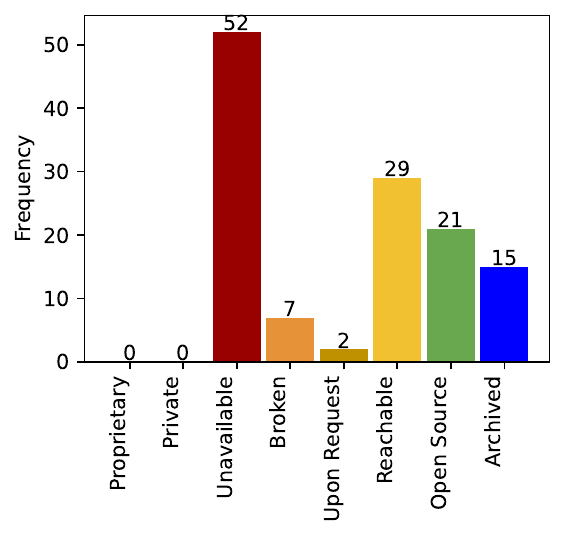}
        \caption{Distribution of availability codes}
        \label{fig:availability:codes}
    \end{subfigure}
    \begin{subfigure}[T]{0.49\textwidth}
        \includegraphics[width=\textwidth]{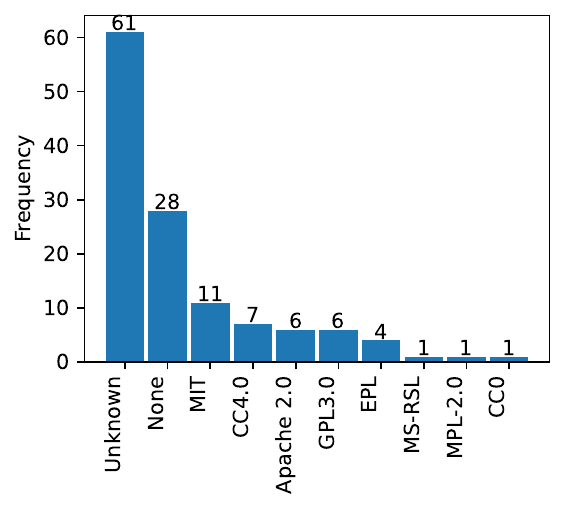}
        \caption{Distribution of used licenses}
        \label{fig:availability:licenses}
    \end{subfigure}
    \caption{Visualization of tool availability}
    \label{fig:availability}
\end{figure}

\Cref{fig:availability} visualizes the status of availability of the extracted tools.
\Cref{fig:availability:codes} shows the distribution of availability codes as per the ranking introduced in \Cref{tab:availability}.
\Cref{fig:availability:licenses} summarizes the licenses used by the extracted tools as well as their frequency in our sample.
Note that the 61 \textit{unknown} licenses are caused by 52 tools being \textit{unavailable} in addition to the 7 tools where the source link was \textit{broken} and the 2 tools that are only available \textit{upon request}.
The 29 tools that were categorized as \textit{reachable} consist of the 28 tools with license type \textit{none} plus one tool published under the Microsoft Reference Source License (MS-RSL)~\cite{gropler2022automated}, which does not count as open source.

\subsection{Discussion}
\label{sec:overview:discussion}

\subsubsection{Interpretation}

The number of NLP4RE tools contained in publications appearing in the last five years of the 12 most prominent venues featuring NLP4RE research speaks for an active field of research. 
Note, however, that this number (126 tools) cannot be directly compared to the 130 tools extracted by Zhao et al.~\cite{zhao2021natural} in the time frame of 1983-2019, since the sampling and inclusion strategy differed.

The distribution of implemented task types to support various RE activities shows clear focal points in NLP4RE research.
Extraction and classification tools for requirements elicitation, classification tools for requirements analysis, and tracing and relating tools for requirements management are favored combinations. 
On the other hand, fundamental algorithms that apply NLP for the particular use in RE but can be used to support multiple activities like zero-shot learning~\cite{alhoshan2023zero} are rare.
Additionally, tools that perform a search and retrieval or generation tasks barely appear in our sample.
The recent rise of large-language models (LLMs) in the broader scope of generative artificial intelligence bear the opportunity to change this.
First applications of LLMs to requirements elicitation~\cite{ronanki2023investigating} and inconsistency detection~\cite{fantechi2023inconsistency} fall outside of the sampled primary studies due to recency, but show first promising results.

The availability of tools as visualized in \Cref{fig:availability:codes} shows a more positive trend towards open source and scientific archival than previous studies~\cite{frattini2023stop}.
The commendable availability of a large portion of the tools ($29+21+15=65$ tools (51.6\%) have an availability status \textit{reachable} or above) correlates with their recency, as the likelihood of authors publishing their artifacts has increased over time~\cite{minocher2020reproducibility} and the availability of non-archived artifacts deteriorates with age~\cite{winter2022retrospective}.
However, the overview shows two remaining shortcomings regarding the availability of NLP4RE tools.
Firstly, almost half of the available tools (availability level \textit{reachable}, $29/65\approx44.6\%$) have no license specified, which impedes their reuse~\cite{rosen2005open}.
Secondly, only $15/65\approx23.1\%$ of the available tools are properly archived, which risks their persistence~\cite{winter2022retrospective}.
The guidelines concluding this chapter, \Cref{sec:management}, aim to provide practical guidance to overcome these shortcomings.

\subsubsection{Challenges and Limitations}

The extracted tools exhibit different levels of maturity.
The maturity of NLP4RE tools ranges from repositories simply containing the necessary scripts and data for replicating a study, like the NFR classification tool\footnote{\url{https://github.com/rgnana1/NFR_Classification_RNN_LSTM}} from Gnanasekaran et al.~\cite{gnanasekaran2021using}, up to well-maintained and -documented tools like the dependency detection tool\footnote{\url{https://github.com/OpenReqEU/dependency-detection}} from the OpenReq initiative by Motger et al.~\cite{motger2019openreq} or the ConDec Jira Plugin\footnote{\url{https://github.com/cures-hub/cures-condec-jira}} by Kleebaum et al.~\cite{kleebaum2021continuous}.
The maturity of tools has a significant impact on their usability but was not analyzed during this study.
Assessing the technology readiness level~\cite{mankins1995technology} of the extracted tools may provide further insights on reusability.

The extraction of the \textit{RE activity} and tool's \textit{task} type categories posed challenges for several tools.
In some cases, the terms used by the authors of a publication did not match the definitions of the types of these categories.
For example, Sainani et al. labeled the task type of their tool as \textit{extraction} even though the actual NLP task performed by the tool is a \textit{classification}~\cite{sainani2020extracting}.
In these cases, we opted for the most appropriate term according to our classification scheme to maintain consistency within our sample.
In other cases, tools employed multiple task types.
For example, Kifetew et al. present a tool which both associates user feedback with system features (task type \textit{tracing \& relating}) and then evaluates the sentiment of the user feedback (task type \textit{extraction})~\cite{kifetew2021automating}.
In these cases, we recorded only the most dominant task type.
However, we reported all of the encountered challenges in a separate sheet contained in our replication package to support the potential extension of our extraction.

\section{Management}
\label{sec:management}

The final section of the chapter provides guidelines aiding the development and management of NLP4RE tools, mainly focusing on two stages of the life cycle of tool development.
\Cref{sec:management:developing} discusses the development and documentation of new and the contribution to existing tools.
\Cref{sec:management:dissemination} provides guidance on disseminating tools, including licensing, version control, and reporting.

\subsection{Development and Documentation}
\label{sec:management:developing}

Developing and documenting NLP4RE tools in a way that supports maintenance, reuse, and improvement is challenging, but essential to ensure a tool provides value to the NLP4RE community. 

\subsubsection{Tool Development}
\label{sec:management:development:dev}

Prior to developing a new NLP4RE tool, an overview of existing tools---for example, the initial list of NLP4RE tools provided by Zhao et al.~\cite{zhao2021natural} and our updated list\textsuperscript{\ref{foot:awesome-list}}---should be consulted to determine if a similar tool has already been proposed and disclosed.
For this, classify the intended tool according to the categories presented in \Cref{sec:classification:task} and identify the subset of tools in the same class.
Review the publications presenting these tools as well as their source---if available---to identify potential candidates that either fully cover the intended goal of the tool, that could be tailored towards it, or that contribute one step of the necessary pipeline.
In case an intended tool does not yet exist, selecting the right technology for implementing a tool becomes relevant.
\Cref{tab:nlp-resources} lists potent libraries and frameworks for different programming languages to consider.
Even more NLP resources for different programming languages can be found at \url{https://github.com/keon/awesome-nlp}.

\begin{table}[hbt!]
    \centering
    \caption{NLP resources for different programming languages}
    \label{tab:nlp-resources}
    \begin{tabularx}{\textwidth}{l|l|X}
        \toprule
        \textbf{Language} & \textbf{Resource} & \textbf{Link} \\ \midrule
        Java & Apache OpenNLP & \url{https://opennlp.apache.org/}) \\
        & Apache UIMA & \url{https://uima.apache.org/} \\
        & Stanford CoreNLP & \url{https://stanfordnlp.github.io/CoreNLP/} \\
        & DKPro & \url{https://dkpro.github.io/} \\ 
        & GATE & \url{https://gate.ac.uk/} \\ \hline
        Python & NLTK & \url{https://www.nltk.org/} \\
        & spaCy & \url{https://spacy.io/} \\
        & HuggingFace & \url{https://huggingface.co/models} \\
        & LangChain & \url{https://python.langchain.com/} \\ \hline
        Javascript & LangChain & \url{https://js.langchain.com/} \\
        \bottomrule
    \end{tabularx}
\end{table}

For some tool types, reference architectures guiding the development of new tools exist.
For example, Dabrowksi et al. have analyzed user feedback miners for app store reviews, a popular NLP4RE tool type, and distilled a reference architecture from the elicited problem domain~\cite{dkabrowski2022mining}.
Empirically derived reference architectures provide great guidance during tool development and should be carefully considered when designing a new tool.

Furthermore, training data needs to be prepared as the quality of every NLP tool depends on the quality of its data.
In case it is impossible to procure data from industry partners, sources like Kaggle,\footnote{\url{https://www.kaggle.com/datasets}} HuggingFace datasets,\footnote{\url{https://huggingface.co/datasets}} or the requirements quality factor ontology\footnote{\url{www.reqfactoront.com}}~\cite{frattini2022live} are valuable starting points.
Another valuable resource is the component-oriented synthetic textural requirements generator CORG~\cite{zaki2021corg}, a tool that ``automatically generate[s] comprehensive combinations of structurally-diverse synthesised textual requirements''~\cite{zaki2021corg} from a dictionary of domain words and verb frames.\footnote{\url{https://github.com/ABC-7/CORG}}

\subsubsection{Documentation}
\label{sec:management:development:doc}

An NLP4RE tool---as any piece of software---ought to be well documented to allow its user to understand, use, and evolve it.
At the minimum, this requires a \texttt{README.md} file located in the top-most directory covering the following sections~\cite{montgomery2023guideline}:

\begin{enumerate}
    \item \textbf{Summary}: a one-paragraph summary of the purpose and objective of the tool
    \item \textbf{Contributor details}: explanation of the ownership and contribution, and information on how to reach out to the contributors
    \item \textbf{Description}: a detailed description of the components of the tool
    \item \textbf{Setup}: a concise and self-contained description on how to set the tool up and use it
    \item \textbf{Licenses}: chosen license for the work (more in \Cref{sec:management:dissemination})
\end{enumerate}

Visualizations aid communicating the developed NLP4RE tool.
Platforms like draw.io\footnote{\url{https://draw.io}} or yEd\footnote{\url{https://www.yworks.com/products/yed}} make it easy to produce diagrams and visualizations summarizing high-level overviews of the tool.

The \textit{setup} section of the \texttt{README.md} file requires special attention for NLP4RE tools.
As most NLP4RE tools rely on third-party technology like the libraries mentioned in \Cref{tab:nlp-resources}~\cite{zhao2021natural}, the setup instruction should contain (1) which resources were used, (2) which versions of these resources are required, and (3) how to install them.
Additional files like a \texttt{requirements.txt} file listing required Python packages and their versions or a Dockerfile\footnote{\url{https://www.docker.com/}} containerizing the setup greatly benefit the long-term usability of any NLP4RE tool.

\subsection{Dissemination}
\label{sec:management:dissemination}

Finally, the successful development of viable NLP4RE tools warrants an equally successful dissemination to share the contribution with the NLP4RE research community and relevant industry sectors. 
This includes hosting the tool in an accessible manner, proper licensing, and thorough reporting.

\subsubsection{Hosting}
\label{sec:management:dissemination:hosting}

NLP4RE tools presented in publications should be made available online to allow replication and evolution~\cite{frattini2023artifacts}.
While disclosing pre-compiled executables makes a tool (re-) usable, disclosing its source code allows further investigation and evolution~\cite{rosen2005open}.
The choice of platform affects the availability of a tool~\cite{frattini2023artifacts}.
Hosting tools on private or institutional platforms has shown to negatively affect their longevity, as these websites are often abandoned when researchers change affiliation~\cite{frattini2023artifacts,gabelica2022many}.
Hence, we recommend using platforms like GitHub\footnote{\url{https://github.com/}} for version-controlled hosting of NLP4RE tools.
The collaborative features implemented in GitHub allow streamlined and effective contribution to repositories of code~\cite{dabbish2012social}.
However, hosting an NLP4RE tool in a GitHub repository still has the following three shortcomings:

\begin{enumerate}
    \item If the owner of the repository renames it, links to the repository (e.g., from the manuscript presenting the tool) will not resolve anymore.
    \item GitHub is currently owned by Microsoft, a for-profit organization not dedicated to the permanence of its content. The company can chose to disable GitHub for the public at any time, making hosted tools unavailable.
    \item Repositories are not located through a digital object identifier (DOI).
\end{enumerate}

Consequently, authors should aim to host their tools on platforms like Zenodo\footnote{\url{https://zenodo.org/}} to ensure permanent, open hosting referenced with an unchanging DOI.
Since Zenodo does not offer the same collaborative features as GitHub, the following synergy ensures utilizing the best of both platforms:
A Zenodo account can be connected to a GitHub account.\footnote{See a detailed guidelines at \url{https://docs.github.com/en/repositories/archiving-a-github-repository/referencing-and-citing-content}}
Then, every \textit{Release} created in the GitHub repository will automatically be archived on Zenodo.
This way, code is hosted in GitHub, making use of its collaborative and version control features, while every milestone is additionally archived on Zenodo, ensuring its permanent accessibility without any additional effort.
For example, the CiRA tool~\cite{fischbach2021cira,frattini2023cira} is hosted on GitHub (\url{https://github.com/JulianFrattini/cira}) and archived on Zenodo (\url{https://zenodo.org/record/8033347}). 
Links between the two (a badge in the Readme linking from the GitHub repository to the Zenodo archive and the automatic backlink from the Zenodo archive to the GitHub repository) allow a seamless transition between the two.

\subsubsection{Licensing}
\label{sec:management:dissemination:licensing}

A tool requires a license as a legal mechanism to make explicit which rights the \textit{licensor} (the owner of the tool according to the license) grants the \textit{licensee} (a user of the tool)~\cite{german2009license}.
Without a license, an NLP4RE tool falls under intellectual property laws and becomes unusable for anyone outside of its owner, even if the code is accessible~\cite{rosen2005open}.
Other forms of legal protection of the tool---like trademarks, trade secrets, or patents~\cite{german2009license}---are also possible, but significantly inhibit the evolution and use of the tool.

\paragraph{Choosing a license}
One important type of license for academic software is the \textit{open source} license.
A license is considered open source if it complies to the criteria defined by the Open Source Initiative\footnote{\url{https://opensource.org/}} listed at \url{https://opensource.org/osd/}.
In short, open source licenses circumvent the restrictions of intellectual property law by explicitly granting a licensee to access, use, and evolve the tool.
A repository containing an open source license is considered open source software~\cite{german2009license}.
The property of open source is imperative for academic software to allow other scholars to learn from it, replicate results, and continue its development~\cite{rosen2005open}.

Hence, we strongly recommend choosing an open source license when disclosing an NLP4RE tool.
Popular open source licenses include the MIT license\footnote{\url{https://opensource.org/license/mit/}} and the GNU General Public License.\footnote{\url{https://opensource.org/license/gpl-3-0/}}
A list of open source licenses can be found at \url{https://opensource.org/licenses/}.
Once you have chosen an appropriate license for your tool, add (1) the raw license text to the top-level directory of the repository and (2) a copyright notice to the bottom of the repository's Readme file.
The copyright notice typically takes the following form: ``Copyright \textcopyright \{year\} \{owner\}. This work is licensed under \{chosen license\}.''
See \url{https://github.com/JulianFrattini/cira} for an example.

\paragraph{Licensing other resources}
Integrating external resources, i.e., code, models, data, etc. that are not owned by the developer of a tool, into an NLP4RE tool requires specific care when hosting the tool.
When using external resources and including them in the repository, the licensee need to make sure that the copyrights of each integrated component allow this reuse and re-publishing~\cite{german2009license}.
If the licenses of external resources allow this, additionally:

\begin{enumerate}
    \item include the raw license text in the same directory as the licensed resource;
    \item include a copyright notice stating who the owner of this resource is next to the licensed resource as a text file;
    \item adjust the copyright notice at the bottom of your own Readme file to list all external resources which are excluded from your tool's license and instead governed by their own license.
\end{enumerate}

A simpler alternative is to not include external resources in a repository, only self-owned material, and instead provide to a user detailed descriptions on how to obtain and integrate these external resources in the Readme file.
This shifts the responsibility of using the external resources from the repository owner to the repository user.
However, this significantly reduces the usability of the tool as it adds additional steps to its setup.

\subsubsection{Reporting}
\label{sec:management:dissemination:reporting}

Finally, reporting an NLP4RE tool in a research article should follow certain guidelines.
To describe the tool, we recommend reporting at least the attributes presented in \Cref{sec:classification}.
Further attributes to report can be derived from more detailed classification schemes, like the NLP4RE ID card~\cite{abualhaija2023replication}.
To reference the tool, we recommend providing a link to both the GitHub repository containing the source code as well as to the Zenodo archive.
In case you need to prioritize, chose the latter, as the link to the Zenodo archive is permanent and contains a reference back to the GitHub repository.

Reporting an NLP4RE tool also includes reporting its evaluation.
Using benchmark tests~\cite{becker2020modern} and adhering to evaluation guidelines like the ECSER pipeline for classifiers~\cite{dell2023evaluating} ensures that the reported results are both complete and comparable, which eases replication and synthesis with other results~\cite{abualhaija2023replication}.

During evaluation, probabilistic components of NLP4RE tools (like classifiers) require to make a trade-off between precision (minimizing the amount of false positives) and recall (minimizing the amount of false negatives).
Deciding which to optimize depends on the context, needs, and expectations of the addressed problem, as well as the cost and likelihood of false positives and false negatives.
A tool with less than 100\% recall will always require a second round of manual reviews complementing the output of the tool~\cite{berry2017evaluation}.
Tools with low precision burden the reader of its output with many irrelevant results (false positives) which introduces skepticism and ultimately rejection of the tool~\cite{femmer2018requirements}.
A conscious choice about which of the two to prefer should be made and documented in the manuscript presenting the evaluation of an NLP4RE tool~\cite{berry2021empirical}.

\section{Conclusion}
\label{sec:conclusion}

This chapter introduces NLP4RE tools by defining a classification scheme (\Cref{sec:classification}), providing an overview over existing tools (\Cref{sec:overview}), and establishing tool management guidelines throughout its life cycle (\Cref{sec:management}).
We hope to incentivize not only the reuse of existing, but also the contribution of new and properly developed, documented, hosted, and maintained tools to the NLP4RE landscape.
The great potential of NLP4RE tools for both academia and industry warrants increasing attention~\cite{dalpiaz2018natural,gorschek2006model}.

Future research avenues do not only include developing new and evolving existing tools, but also establishing a benchmark infrastructure specifically for the NLP4RE research domain.
While NLP benchmarks exist~\cite{becker2020modern} and contain popular resources like SQuAD\footnote{\url{https://rajpurkar.github.io/SQuAD-explorer/}}~\cite{rajpurkar2018know} and GLUE\footnote{\url{https://super.gluebenchmark.com/}}~\cite{wang2019superglue}, the NLP4RE domain lacks any specific benchmark infrastructure.
Investing effort into NLP4RE-specific benchmark data sets as well as an infrastructure to track and compare tool performances will improve the synthesis of results~\cite{abualhaija2023replication} and advance progress in the NLP4RE tool development field further~\cite{dalpiaz2018natural}.

\bibliographystyle{splncs04}
\bibliography{material/references}


\end{document}